\documentclass{iaus}
\usepackage{graphicx}

\title[Star Formation Thresholds] 
{Star Formation Thresholds}

\author[J. Schaye]   
{Joop Schaye$^1$}

\affiliation{$^1$Leiden Observatory, Leiden University, P.O. Box 9513,
  2300 RA Leiden, the Netherlands \break email: schaye@strw.leidenuniv.nl}

\pubyear{2007}
\volume{244}  
\pagerange{299--307}
\date{?? and in revised form ??}
\jname{Dark Galaxies \& Lost Baryons}
\editors{J. I. Davies \& M. J. Disney, eds.}

\newcommand{\apj}{\textit{ApJ}}
\newcommand{\apjl}{\textit{ApJ}}
\newcommand{\mnras}{\textit{MNRAS}}
\newcommand{\aj}{\textit{AJ}}

\newcommand{\Msolpcsq}{\mbox{M}_{\odot}\,\mbox{pc}^{-2}}

\newcommand{\K}{\mbox{K}}

\newcommand{\ion}[2]{\hbox{#1\,{\sc #2}}}
\newcommand{\HI}{\ion{H}{I}} 
\newcommand{\HII}{\ion{H}{II}} 

\newcommand{\cm}{{\rm cm}}

\newcommand{\s}{{\rm s}}
\newcommand{\kms}{{\rm km}\,{\rm s}^{-1}}
\newcommand{\kpc}{{\rm kpc}}
\newcommand{\pc}{{\rm pc}}

\newcommand{\Msun}{{{\rm M}_\odot}}

\newcommand{\la}{\mathrel{\mathchoice{\vcenter{\offinterlineskip\halign{\hfil
$\displaystyle##$\hfil\cr<\cr\sim\cr}}}
{\vcenter{\offinterlineskip\halign{\hfil$\textstyle##$\hfil\cr<\cr\sim\cr}}}
{\vcenter{\offinterlineskip\halign{\hfil$\scriptstyle##$\hfil\cr<\cr\sim\cr}}}
{\vcenter{\offinterlineskip\halign{\hfil$\scriptscriptstyle##$\hfil\cr<\cr\sim\cr}}}}}

\begin{document}

\maketitle

\begin{abstract}
To make predictions for the existence of ``dark galaxies'', it is necessary
to understand what 
determines whether a gas cloud will form stars. Star formation
thresholds are generally explained in terms of the Toomre criterion
for gravitational instability. I contrast this theory with the
thermo-gravitational instability hypothesis of Schaye (2004), in which
star formation is triggered by the formation of a cold gas phase and
which predicts a nearly constant surface density threshold. 
I argue that although the Toomre analysis is useful for the
global stability of disc galaxies, it relies on assumptions that break
down in the outer regions, where star formation thresholds are
observed. The thermo-gravitational instability hypothesis can account
for a number of observed phenomena, some of which were thought to
be unrelated to star formation thresholds.
\keywords{stars: formation, ISM: structure, galaxies: formation,
  galaxies: ISM}
\end{abstract}

\firstsection 
\section{Introduction}

Do all gas clouds with $M \gg M_\odot$ contain stars? And if not, why do
some clouds form stars whereas other do not? Observationally, it is
nearly impossible to prove that a gas cloud has no stars at all,
because a small number of old stars would generally be far too faint to
detect. While it is difficult to put upper limits on the amount
stellar mass, it is relatively easy to obtain interesting upper limits
on the amount of star formation. This can for example be done through the
non-detection of recombination lines such as H$\alpha$, which are
generated in the \HII\ regions around young stars, or through infrared
observations.  

The \HI\ components of most spiral galaxies are observed to extend
beyond the region of active star formation. It has been known for
some time, that star formation is generally limited to regions where the total gas
surface density exceeds $\Sigma_{\rm c}\sim 10~\Msolpcsq$ (e.g., Guiderdoni
1987; Skillman 1987, Taylor et al.\ 1994).
The existence of a surface density threshold for star
formation is, however, usually explained in terms of the Toomre
criterion for gravitational instability (e.g., Quirk 1972; Kennicutt 1989). 
Neither rotation nor pressure can stabilize a thin, gaseous,
differentially rotating disk if the Toomre $Q$ parameter, 
\begin{equation}
Q \equiv {c_{\rm s} \kappa \over \pi G \Sigma_{\rm g}},
\label{eq:Q}
\end{equation}
is less than unity (Safronov 1960; Toomre 1964); where the effective sound
speed $c_{\rm s}$, the epicycle frequency $\kappa$, and the gas surface
density $\Sigma_{\rm g}$ all depend on radius $r$. According to this
theory, star formation is suppressed when the gas surface density falls below the
critical value
\begin{equation}
\Sigma_{\rm c} \equiv {c_{\rm s} \kappa \over \pi G} \equiv Q\Sigma_{\rm g}
\end{equation}
which depends on the local values of the velocity dispersion and the
epicycle frequency.

Kennicutt (1989) and Martin \& Kennicutt (2001, hereafter MK01) tested
the hypothesis that the Toomre criterion is responsible for the observed
cutoff in the star formation rate by measuring the ratio $\alpha
\equiv 1/Q$ of
the gas surface density $\Sigma_{\rm g}$ to the critical surface density.
Assuming a constant velocity
dispersion of $6~\kms$ and after azimuthal averaging, MK01
found that $\alpha \equiv \Sigma_{\rm g} / 
\Sigma_{\rm c} \approx 0.5$\footnote{We corrected their value of $\alpha$,
  see Schaye (2004).} at the truncation radius in a sample of 32 nearby spiral
galaxies. 

Elmegreen \& Parravano (1994)
emphasized that the need for a cold gas phase imposes a pressure
threshold. Schaye (2004, hereafter S04) showed that the transition
from the warm to 
the cold gas phase triggers gravitational instability on a wide range
of scales. He demonstrated that this thermo-gravitational instability
can account for the observed star formation threshold, which he
predicted to fall in the range $\Sigma_{\rm c} \approx 3$--$10~\Msolpcsq$
(corresponding to a critical pressure $P_c/k \sim
10^2$--$10^3~\cm^{-3}\,\K$ and a critical volume density $n_{\rm H}
\sim 10^{-2}$--$10^{-1}~\cm^{-3}$), with weak dependencies on the
metallicity, UV radiation field, turbulent pressure, and the mass
fraction in gas.

After briefly summarizing the S04 model in section \ref{sec:s04model}, 
I argue in section \ref{sec:toomreassumptions} that the Toomre
$Q$ hypothesis relies on assumptions that break down in the outer
parts of galaxies, where star formation 
thresholds are observed. In section \ref{sec:obs} I discuss a number of
observational findings that are naturally explained by the
thermo-gravitational instability hypothesis. 

In the second part of my talk, I discussed the relation between star
formation laws expressed in terms of surface densities, volume densities and
pressures. I presented a method based on this theory with which it is
possible to reproduce arbitrary Kennicutt-Schmidt laws using numerical
simulations, without any free parameters. I do not have the space here
to describe this model and its implications for the interpretation of
simulations. For more on this topic, stay tuned for the paper ``On the
relation between the Schmidt and Kennicutt-Schmidt star formation laws
and its implication for numerical simulations'' (Schaye \& Dalla
Vecchia, in preparation). 

\section{Thermo-gravitational instability}
\label{sec:s04model}

The insterstellar medium of galaxies is highly complex. It contains a
number of 
gas phases, including a cold ($T \la 10^2~\K$), warm ($T\sim 10^4~\K$),
and hot ($T\gg 10^4~\K$) phase. Feedback from star formation drives
strong, supersonic turbulence which can both promote and suppress
gravitational stability. Because of this complexity, it is difficult
to predict exactly where and when star formation occurs. However, the
outer parts of galaxies are much simpler. Here the gas is almost
exclusively warm and there is no, or very little, star formation and hence
turbulence is weak and typically subsonic. Moving in towards the
center, we may therefore be able to predict where star formation
begins, even if we cannot predict what happens further in.

S04 noted that the UV background radiation implies the existence of a
threshold surface density for the formation of a cold phase, which
agrees with the empirically determined star formation threshold. He
hypothesized that the drop in the pressure associated with the
transition to a cold phase triggers gravitational instability and thus
star formation. Conversely, for densities below the threshold,
self-gravitating gas clouds are kept warm and stable by the UV
radiation. 

To test this hypothesis, S04 constructed two model
galaxies consisting of exponential discs 
embedded in cold dark matter halos. The discs were illuminated by UV
radiation and contained metals and dust. For a fixed surface density,
the temperature and volume density were determined numerically under
the condition of vertical hydrostatic equilibrium. The radiative
transfer and chemistry were calculated using the publicly available
package \textsc{CLOUDY} (Ferland 2000). 

\begin{figure}
\resizebox{\textwidth}{!}{\includegraphics{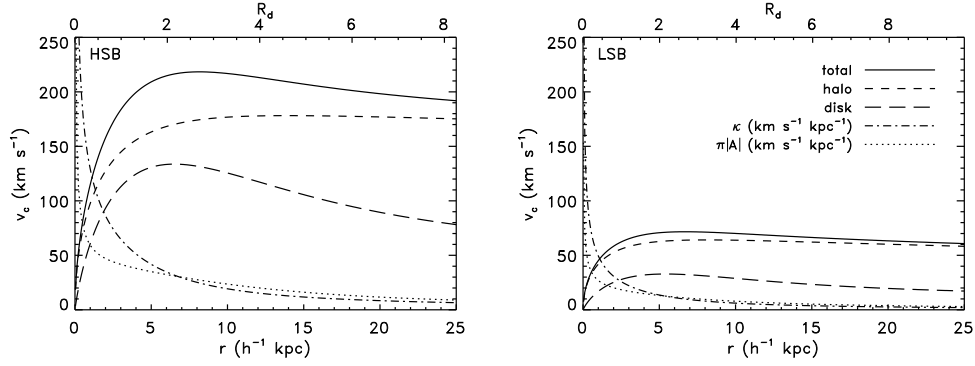}}
  \caption{Circular velocity as a function of radius for
models HSB (left) and LSB (right). The short (long) dashed curve shows the
contributions of the halo (disc). Also shown are the epicycle frequency
(dot-dashed curves) and $\pi$ times Oort's constant $A$
(dotted curves), both in $\kms\,\kpc^{-1}$. Figure taken from
S04.}
\label{fig:rotcurves}
\end{figure}

The parameters of the model galaxies were chosen to mimic a high and a
low surface brightness galaxy. Figure~\ref{fig:rotcurves} shows the
rotation curves (solid), including the contributions of the halo
(short dashed) and disc (long dashed), the epicycle frequency
(dot-dashed) and $\pi$ times Oort's constant
(dotted). 

\begin{figure}
 \resizebox{\textwidth}{!}{\includegraphics{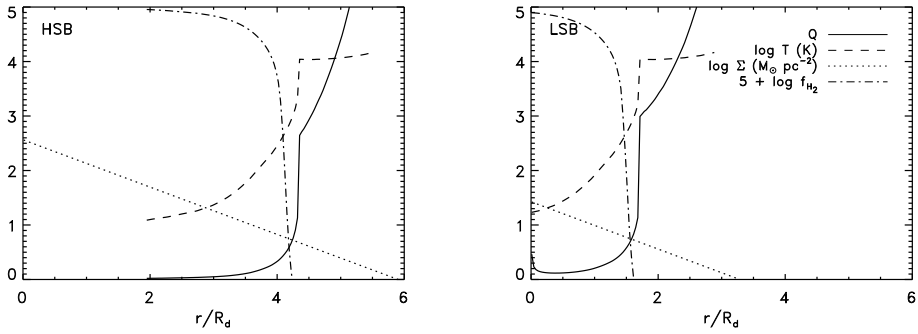}}
  \caption{The Toomre $Q$ parameter (solid curves), the
temperature $\log T$ (dashed curves), the molecular fraction
$5+\log f_{H_2}$ 
(dot-dashed curves) and the surface density $\log\Sigma (\Msun\,\pc^{-2})$
(dotted curves) are all plotted as a function of radius 
for models HSB (left) and LSB (right). 
Figure taken from S04.\label{fig:threshold}}
\end{figure}

The solid curves in Figure~\ref{fig:threshold} show the
Toomre $Q$ parameter as a function of radius. The sudden drop in
the $Q$ values coincides with (and is caused by) a similar drop in the
temperature (dashed curves), and a sharp increase in the molecular
fraction (dot-dashed curves). The
transition to the cold phase, which coincides with 
the onset of gravitational instability ($Q<1$), occurs at a fixed
surface density. Note that the models become unrealistic shortwards of
the critical radius, where feedback from star formation will increase
the UV field, the metallicity, and the turbulent pressure. 

\begin{figure}
 \resizebox{\textwidth}{!}{\includegraphics{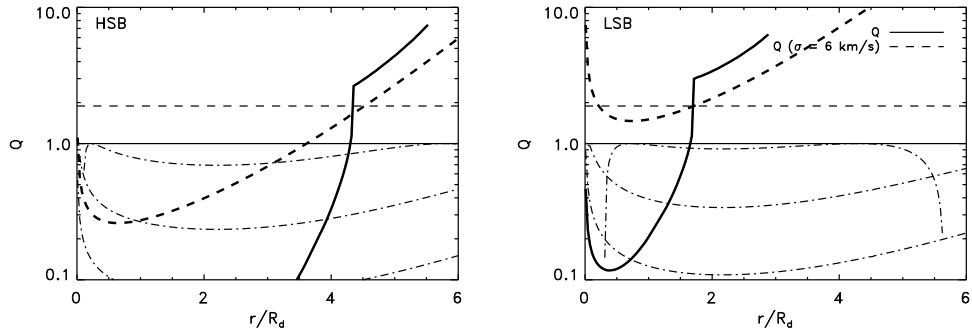}}
  \caption{The Toomre $Q$ parameter as a function of radius
(thick solid curves) for models HSB (left) and LSB
  (right). The solid horizontal lines indicate $Q=1$. 
The dot-dashed curves
indicate the $Q$ values below which the disc is unstable to
perturbations with wavelengths of (top) $10^3$, (middle)
$10^2$, and (bottom) 10~pc. The thick dashed curves show the Toomre
$Q$ parameter computed under the assumption that $\sigma=6~\kms$. The
horizontal dashed lines indicate the critical $Q$ value measured by
Kennicutt (1989) under the assumption that $\sigma=6~\kms$. Figure taken
from S04.\label{fig:Qcomparison}} 
\end{figure}

Figure~\ref{fig:Qcomparison} again shows the $Q$ parameter (thick
solid curves) as a function of radius, but note that this time the
$y$-axis is logarithmic. The value $Q=1$ is indicated with solid, 
horizontal lines. A Toomre $Q$ of unity implies
marginal instability only for fluctuations with $\lambda=\lambda_{\rm
  crit}/2$, where
\begin{equation}
\lambda_{\rm crit} \equiv {4\pi^2 G \Sigma_{\rm g} \over \kappa^2}. 
\end{equation}
In general, instability of perturbations on scale $\lambda$ requires
$Q<Q_c(\lambda)$, where
\begin{equation}
Q_c(\lambda) = 2\sqrt{{\lambda \over \lambda_{\rm crit}} - \left
  ({\lambda \over \lambda_{\rm crit}}\right )^2}.
\end{equation}
The dot-dashed curves in Figure~\ref{fig:Qcomparison} indicate the $Q$
values below which the disc is unstable to
perturbations with wavelengths of (top) $10^3$, (middle)
$10^2$, and (bottom) 10~pc. The transition to the cold phase causes
$Q$ to drop far below unity, thereby triggering instability on a wide
range of scales. 

Assuming a constant velocity dispersion of $\sigma =
6~\kms$ (thick dashed curves) would result in large errors in the
$Q$-parameter. However, if we recalibrate the critical $Q$ parameter
and compare with the value measured by Kennicutt (1989) (horizontal
dashed lines), then we obtain agreement between the predicted critical
radii (the intersections of the thick solid curves and the horizontal
solid lines roughly coincide with the intersections of the thick
dashed curves and the horizontal dashed lines).

S04 computed a large range of models and found that the following
function can describe the variation of the critical surface density with
the parameters of the model
\begin{equation} 
\log\left ( {\Sigma_{\rm c} \over 1 \Msolpcsq} \right )\approx 0.8 + 0.3\log(f) - 0.3
\log\left ({Z \over 0.1 Z_\odot}\right )  + 0.2 \log\left ({I \over
  10^6~\cm^{-2}\,\s^{-1}}\right ),  
\label{eq:sigmac_s04}
\end{equation}
where $Z$ is the metallicity, $I$ is the flux of ionizing radiation,
and $f\equiv f_g/f_{\rm th}$, with $f_{\rm g}$ the mass fraction in gas and
$f_{\rm th}$ the fraction of the pressure that is thermal. The above
formula (see S04 for second order corrections) was found to reproduce
the results for a wide variety of galaxy models.

\section{The assumptions underlying the Toomre hypothesis are invalid
  in outer discs}
\label{sec:toomreassumptions}
In this section I will argue that the Toomre analysis is generally not
applicable to star formation thresholds in the outer parts of disc galaxies,
because the underlying assumptions are invalid. Below I will discuss
the offending assumptions in turn.

\begin{description}

\item[A Toomre $Q$ smaller than unity is sufficient for star formation]~
A Toomre $Q$ of unity implies
marginal instability only for fluctuations with $\lambda=\lambda_{\rm
  crit}/2$. In the outer parts of disc galaxies, this critical wavelength is
typically $> 1~\kpc$ (see Fig.~\ref{fig:Qcomparison}). Hence, while
$Q=1$ is relevant for large-scale 
modes such as spiral arms, we require fluctuations with much smaller
wavelengths to become unstable to account for the formation of star
clusters. Note that 
if $\Sigma_{\rm g} \sim \Sigma_{\rm c}(\sigma=6~\kms)$ is observed, 
then the condition $Q\ll 1$ implies $\Sigma_{\rm c} \ll
\Sigma_{\rm c}(\sigma=6~\kms)$ and thus $\sigma \ll 6~\kms$.  

\item[The disc is thin]~
In the outer parts of galaxies, the disc scale height is similar to 
the most unstable wavelength, thereby invalidating the assumption that
the disc is thin.

\item[The perturbations are axisymmetric]~
Beyond the truncation radius identified in azimuthally averaged
observations, star formation does occur in isolated regions whenever
the local gas surface density exceeds the global 
threshold (e.g., Ferguson et al.\ 1998). The Toomre analysis is valid
only for axisymmetric 
perturbations and can therefore not straightforwardly be applied to
such regions. 

\item[The velocity dispersion is constant]~
Although the Toomre analysis does not require it, when it is applied
to star formation thresholds it is generally assumed that the velocity
dispersion 
is constant. Following Kennicutt (1989), most work assumes $\sigma =
6~\kms$. For 
sufficiently large scale heights, the gas will be kept warm by the UV
background radiation. Therefore, there will always be a component with
a velocity dispersion of at least that of the warm phase. However, if
there is any cold gas then the relevant velocity dispersion, and
thus $Q$, can locally be much smaller. 

\item[Rotation needs to be taken into account]~
As mentioned above, to form stars, or even star clusters, we require 
instability for $\lambda \ll \lambda_{\rm
  crit}$ and hence $Q\ll 1$. However, for $\lambda \ll \lambda_{\rm
  crit}$ the Toomre criterion reduces to the two-dimensional Jeans criterion:
$\lambda > c_{\rm s}^2 / G\Sigma_{\rm g}$. Hence, rotation is generally not
directly relevant for star formation thresholds.

The exception is the central part of the disc,
where $\lambda_{\rm crit} \propto \kappa^{-2}$ becomes small (see
Fig.~\ref{fig:rotcurves}) and
the Coriolis force can, in principle, prevent collapse on scales
relevant for star formation. However, in practice the disc cannot be considered
thin in the central parts because the scale height is no longer small
compared with the radius. Moreover, the bulge will need to be taken
into account.

\end{description}
 
\section{Observations that are naturally explained by the
  thermo-gravitational instability}
\label{sec:obs}
A number of observational findings, some of which were thought
to be unrelated, can be explained by the
thermo-gravitational instability theory of star formation thresholds. Below I will
discuss the most important ones.

\subsection{Why is the critical surface density constant?}

Why can a single constant, surface density threshold account for the
observed star formation thresholds (e.g., Guiderdoni 1987; Skillman 1987, Taylor
et al.\ 1994)? A constant threshold works works even
on a pixel-by-pixel
basis for high-resolution observations of nearby galaxies (de Blok \&
Walter 2006) and it can also account for the isolated star formation events beyond
the optical radii of galaxies (e.g., Ferguson et al.\ 1998).

If the Toomre hypothesis holds, then $\Sigma_{\rm c} \propto c_{\rm s}\kappa$ and
should therefore vary from place to place. Even if we fix $c_{\rm s}$, as is
often done, it should still depend on the epicycle frequency. If the
thermo-gravitational instability is responsible, then the critical
surface density will be nearly constant [see equation
  (\ref{eq:sigmac_s04})] and the predicted value is in excellent
agreement with the observations. 

Note that although $\Sigma_{\rm c}$ is predicted to depend only very weakly
on environmental factors such as the metallicity and the UV intensity,
the differences may still become noticeable in extreme
environments. For example, we predict $\Sigma_{\rm c} \gg 10~\Msolpcsq$ for
the first generation of stars at very 
high redshift (because $Z \ll 0.1 Z_\odot$) and also very near
strong sources of UV radiation.

\subsection{Why is a large fraction of disc galaxies inconsistent
  with the Toomre hypothesis?}

Why does a constant threshold work even when the
Toomre criterion does not? Many studies have found so-called
``subcritical discs''. That is, galaxies for which $Q(\sigma =
6~\kms)$ exceeds the critical value throughout the region of active
star formation (e.g., MK01; Wong \& Blitz 2002; Auld et al.\ 2006). About 1 in 4
of the galaxies from the sample of MK01 fall in this category. As can
be seen from the figures in these papers, the subcritical discs do
agree with the constant star formation threshold hypothesis.

\subsection{Why is the velocity dispersion in the outer disc $8~\kms$?}

Why is the velocity dispersion of the outer \HI\ disk constant and why
does it have the value $8~\kms$ (e.g., Lo et al.\ 1993; Meurer et al.\
1996)? Neither magnetically induced (Sellwood \& Balbus 1999) nor
gravitationally induced (e.g., Wada et al.\ 2002) turbulence can
account for such a high (and constant) value. 

According to the thermo-gravitational instability picture, the gas in
the outer disk, at surface densities below the threshold, is kept warm
by the UV background radiation. Although this should be
uncontroversial, theoretical studies have rarely taken radiation into
account.  

\subsection{Why does the same $\Sigma_{\rm c}$ also work for tidal arms?}
Maybhate et al.\ (2007) found that the standard, constant surface
density threshold can also account for observations of star clusters
in tidal arms. Tidal arms can hardly be considered part of thin,
differentially rotating discs, and the Toomre hypothesis is therefore
even less applicable to these observations than to those in the outer
parts of disc galaxies. However, the thermo-gravitational instability
does not assume a thin, rotating disc and should thus hold equally well in
tidal arms. 

\subsection{Why is gas with $\Sigma_{\rm g} < \Sigma_{\rm c}$ nearly always warm?}
Why does the contribution of the cold interstellar phase to the
gas surface density become negligible beyond the edge of the star forming
disk (Braun 1997)? This observation is in excellent agreement with the
predictions of the thermo-gravitational instability hypothesis: the
formation of a cold phase triggers star formation and the two
therefore go hand in hand. For $\Sigma < \Sigma_{\rm c}$ the cold gas
fraction, and thus the star formation rate, plummets. 

In the constant velocity dispersion version of the Toomre hypothesis
there is no 1-1 relationship between the 
ability to form a cold phase and the ability to form stars. While
instability of the warm phase will presumably also eventually trigger
the formation of a cold phase, the converse does not need to be
true. Therefore, there is no reason why there could not be lots of
cold gas in regions where there is no star formation.

\subsection{Why does the H I surface density saturate at $\Sigma_{\rm
    HI} \sim \Sigma_{\rm c}$} Many observations indicate that while the \HI\
surface density increases with decreasing galactocentric radius, it
typically saturates around $10~\Msolpcsq$ (e.g., Wong \& Blitz
2002). At smaller radii the total gas surface density can be much
higher, but for $\Sigma_{\rm g} \gg 10~\Msolpcsq$ most of the gas is
molecular. Is it a coincidence that the \HI\ surface density saturates
at the star formation threshold? If the Toomre hypothesis is correct, then it
would be. Although star formation may destroy the \HI, there will be galaxies for
which the disc is Toomre stable for $\Sigma_{\rm g} \gg 10~\Msolpcsq$. 

In Schaye (2001) I pointed out that there is a physical upper limit to
the \HI\ surface density of gas clouds because the gas will quickly go
molecular beyond a critical density, which depends
weakly on the metallicity and the UV radiation field. The transition
to a cold, molecular phase both limits the \HI\ surface density and
triggers star formation. Thus, the agreement between the critical surface density
for star formation and the \HI\ saturation limit is a prediction of the
thermo-gravitational instability picture.

It has been argued that MK01's finding that the molecular fraction
varies wildly at the critical radius for star formation is evidence against the
thermo-gravitational hypothesis. However, a large scatter in the
molecular fraction is exactly what is expected here. The critical
radius $R_c$ is defined (by eye) as the 
radius within which star formation is ubiquitous and beyond which it is
sporadic. The prediction is that for $R\gg R_c$ the gas is mostly
atomic, while for $R\ll R_c$ it is
mostly molecular (provided that $\Sigma_{\rm g}(r)$ is well behaved so that
$\Sigma_{\rm g} \ll \Sigma_{\rm c}$ for $R\gg R_c$ and $\Sigma_{\rm g} \gg \Sigma_{\rm c}$ for
$R\ll R_c$). Since $R\sim R_c$ is the transition regime, we expect
both atomic and molecular regions. Azimuthal averaging will then
result in a wide range of molecular fractions.

\subsection{Why is the empirically determined critical $Q$ different
  for irregulars?} 

MK01 find that star formation is suppressed when $\Sigma_{\rm g}/\Sigma_{\rm c} > \alpha =
0.5$. On the other hand, Hunter, Elmegreen, \& Baker 
(1998) measured a smaller value of $\alpha$ for a sample of
irregulars. However, Hunter et al.\ (1998) used a velocity dispersion
of $9~\kms$ because the standard value of $6~\kms$ was inconsistent
with the observations. Had they assumed used $6~\kms$, they would have
inferred a value of $\alpha$ that agrees within $1\sigma$ with the
Kennicutt (1989)
value. This again reflects the fact that a constant surface density
threshold describes the observations very well. If $\sigma$ is
increased, then a constant surface density threshold implies a
corresponding reduction of $\alpha$.

\section{Conclusions}

S04 argued that star formation thresholds are set by the necessity to form a cold
gas phase. Below a critical surface density the gas is kept warm by UV
radiation. Conversely, when the surface density exceeds the threshold,
the drop in the gas pressure associated with the phase transition
triggers gravitational instability. As we discussed above, this
hypothesis is in excellent, quantitative agreement with observations
of the outer parts of galaxies. It can account for the observed star formation
thresholds and the observed velocity
dispersion, it can explain the lack of cold
gas beyond regions of active star formation, and it also explains the coincidence
between the critical surface density and the \HI\ saturation limit.

Does the Toomre $Q$ play any role at all? It almost certainly does
play an important role. Although it is not applicable to the outer
parts, the molecular disc is thin compared with $\lambda_{\rm crit}$
and can thus be characterized using the Toomre parameter. If $Q<1$ then the
disc is unstable to large-scale axisymmetric perturbations, which will
for example lead to the formation of spiral arms. The resulting
increase in the surface density can trigger local instabilities, and
not necessarily only through the formation of a cold phase. However,
on scales relevant to the formation of star clusters, the 3-D Jeans
criterion describes the local stability better than the local Toomre
$Q$, and much better than the constant velocity dispersion version of
the Toomre criterion. 

\begin{acknowledgments}
I would like to thank the organizers for what turned out to be a
very stimulating conference. This work
was supported by Marie Curie Excellence Grant MEXT-CT-2004-014112. 
\end{acknowledgments}

\begin{discussion}

\discuss{Taylor}{Recent GALEX observations do not show evidence for star formation
thresholds. How do you interpret those observations?}

\discuss{Schaye}{The GALEX observations are very interesting. At first
sight they appear to contradict the H$\alpha$ observations, which do
show evidence for star formation thresholds. GALEX detects star light, whereas the
H$\alpha$ comes from \HII\ regions. GALEX is sensitive to star formation up to
100~Myr after the stars formed, whereas the H$\alpha$ becomes
difficult to detect after 10~Myr. The latter observations therefore
offer more direct tests of theories of star formation thresholds. It is
difficult to use GALEX observations to relate star formation to local gas
densities because after 100~Myr feedback from star formation may well
have destroyed/removed all the local \HI\ and H$_2$. GALEX may probe the
cumulative effect of the isolated star formation events that are seen in
H$\alpha$ in the outer discs.}

\discuss{Taylor}{Stellar discs may not be truncated, although many do
show evidence for a change in the exponential scale length in the
outer parts. Can star formation thresholds explain this?}

\discuss{Schaye}{Star formation does occur in the far outer parts of
galaxies, when the local gas density exceeds the threshold. I
could imagine some kind of change in the stellar surface density
profile around the point where the azimuthally averaged gas surface
density equals the threshold. Within that critical radius the star formation
rate will be mostly determined by the efficiency of star formation
rather than by the threshold. On the other hand, beyond the critical
radius it will be determined more by the frequency of density
fluctuations strong enough to exceed the threshold. However, the
stellar disc is built up over long time scales, during which the
critical radius may vary. Moreover, stars may move radially after they
formed and stars in the outer discs may also have been accreted. These
effects complicate the relationship between star formation thresholds
and stellar surface density profiles.}

\end{discussion}

\end{document}